# Active colloids in liquid crystals

Oleg D. Lavrentovich

Liquid Crystal Institute and Chemical Physics Interdisciplinary Program

Kent State University, Kent, OH 44242

olavrent@kent.edu

tel 1 330 672 4844

fax 1 330 672 2796

## Abstract

Active colloids in liquid crystals (ACLCs) is an active matter with qualitatively new facets of behavior as compared to active matter that becomes isotropic when relaxed into an equilibrium state. We discuss two classes of ACLCs: (i) "externally driven ACLCs", in which the motion of colloidal particles is powered by an externally applied electric field, and (ii) "internally driven ACLCs", formed by self-propelled particles such as bacteria. The liquid crystal (LC) medium is of a thermotropic type in the first case and lyotropic (water based) in the second case. In the absence of external fields and self-propelled particles, the ACLCs are inactive, with the equilibrium LC state exhibiting long-range orientational order. The external electric field causes ACLCs of type (i) to experience translations, rotations, and orbiting, powered by mechanisms such as LC-enabled electrokinetics, Quincke rotations and entrapment at the defects of LC order. A dense system of Quincke rotators, orbiting along circularly shaped smectic defects, undergoes a transition into a collective coherent orbiting when their activity increases. An example of internally driven ACLCs of type (ii) is living liquid crystals, representing swimming bacteria placed in an otherwise passive lyotropic chromonic LC. The LC strongly affects many aspects of bacterial behavior, most notably by shaping their trajectories. As the concentration of bacteria and their activity increase, the orientational order of living liquid crystals experiences two-stage instability: first, the uniform steady equilibrium director is replaced with a periodic bend deformation, then, at higher activity, pairs of positive and negative disclinations nucleate, separate, and annihilate in dynamic patterns of topological turbulence. The ACLCs are contrasted to their isotropic counterparts.

# 1. INTRODUCTION

This review deals with two broad types of active colloids in liquid crystals (ACLCs), classified according to the driving principle. The first class is "externally driven ACLCs", in which the motion of colloidal particles is powered by an externally applied field, such as the electric field. The second class is an "internally driven ACLCs", formed by self-propelled particles. In the absence of external fields and self-propelled particles, the liquid crystal (LC) is inactive. In the ground state, the LC exhibits a spatially uniform long-range orientational order. The simplest type of LCs, the so-called uniaxial nematic, exhibits a single direction, called the director, $\hat{\mathbf{n}}$, with the properties $\hat{\mathbf{n}}^2 = 1$ and $\hat{\mathbf{n}} \equiv -\hat{\mathbf{n}}$, along which the molecules (or their aggregates) are aligned in space. The ACLCs are thus complementary to the so-called active nematics [1-4] in which dense populations of living or artificial moving particles can exhibit orientational order when driven out of equilibrium.

The anisotropy of LCs caused by their long-range orientational order affects the active colloids in two aspects. First, the director field impacts the trajectories of active particles. In most cases, motion parallel to $\hat{\mathbf{n}}$ is easier than motion perpendicular to $\hat{\mathbf{n}}$; if the director field is non-uniform, the trajectories follow the local direction of $\hat{\mathbf{n}}$. If certain regions show a strong gradient of the LC order, the moving particles can be trapped in them. Second, the orientational order imparts new mechanisms by which the colloidal particles become active and convert the energy of an external field into motion. These mechanisms are rooted in anisotropy of LC properties such as electric conductivity and dielectric permittivity.

The purpose of this review is to present the studies on ACLCs that appeared within the last decade and to highlight the differences between the ACLCs and an "isotropic" version of active colloids, recently reviewed by Aranson [5*] and Degen [6*], in which the medium, when deprived of the active colloidal component, shows no orientational order on its own; water is an example of such a medium. We start with the basic properties of LCs that are essential for understanding the mechanisms of the dynamical behavior of colloids.

## 2. BASIC PROPERTIES OF LIQUID CRYSTALS AND COLLOIDS IN THEM

Because of the orientational order, all physical properties of LCs, such as dielectric permittivity and electric conductivity, are anisotropic. For example, the electric conductivity $\sigma_\parallel$ measured parallel to $\hat{\mathbf{n}}$ is generally higher than the conductivity $\sigma_\perp$ in the perpendicular direction.

Any deviation of the director from the uniform state $\hat{\mathbf{n}}(\mathbf{r}) = \text{const}$ costs some elastic energy, usually described within the Frank-Oseen formalism [7, 8]. If the nematic experiences gradients of the director, on the order of $1/L$, over a length $L$, then the elastic energy scales as $KL$, where $K$ is an average value of the Frank elastic constants, on the order of 10 pN. Anisotropic molecular interactions are also responsible for the anisotropy of surface properties of the LCs by establishing a preferred surface orientation of the director at the boundary, the so-called "easy axis" $\hat{\mathbf{n}}_0$. To deviate the actual director $\hat{\mathbf{n}}$ from $\hat{\mathbf{n}}_0$, one needs to spend some work, the measure $W$ of which is called the surface anchoring strength. Experimental data on $W$ vary broadly, $W \sim (10^{-6} \div 10^{-3})$ J/m$^2$ [8]. The ratio $\lambda = K/W$ is called the de Gennes-Kleman length; with the estimates above, one finds $\lambda = (0.1-10)\,\mu\text{m}$, much larger than the molecular scale $l = (1-10)\,\text{nm}$. The feature leads to interesting observations regarding colloidal particles in a LC. A colloidal particle of radius $R \ll \lambda$ placed in a uniform LC, does not perturb the director. A much larger particle, $R \gg \lambda$, in contrast, causes substantial director distortions over distances comparable to $R$, Fig.1. The reason is that the elastic energy scales as $KR$ while the surface anchoring energy of the particle-LC interface scales as $WR^2$. At $R \approx \lambda$, the elastic and surface anchoring energies are in fine balance; at $R \gg \lambda$, surface anchoring sets the surface director firmly along the preferred directions and thus dictates a strongly distorted equilibrium director field in the neighborhood of the particle; at $R \ll \lambda$, the elastic energy prevails, establishing a fairly uniform director at the expense of large departures of the surface director from the preferred anchoring direction.

The elastic energy associated with the director distortions around particles is enormous; for a micron-size sphere, $F_{elastic} \sim KR \sim 10^4 k_B T$, where $k_B$ is the Boltzmann constant, $T$ is the absolute temperature, corresponding to the range of existence of the LC state. These distortions are responsible for long-range anisotropic interactions of colloids in LCs [9]. Equilibrium structures of LC colloids shaped by the LC elasticity and anchoring have been the subject of intense studies since the work by Poulin et al. [9], as reviewed in Refs. [10-15*]. The balance of anchoring and elastic forces also leads to the effect of colloidal levitation in an LC environment thanks to the elastic repulsion from the bounding wall [16-19].

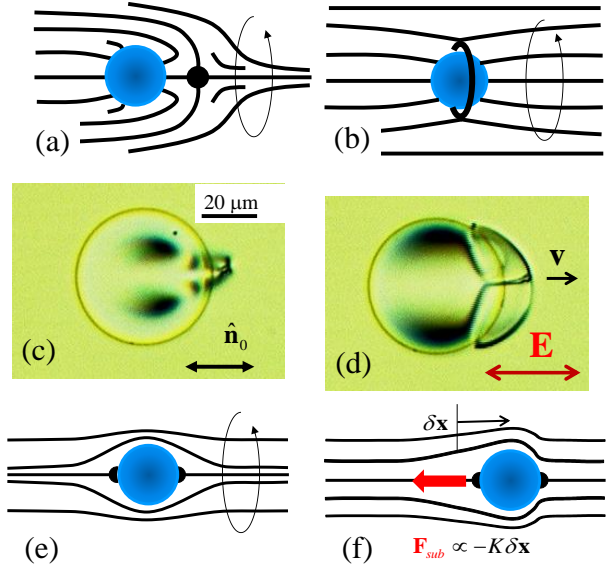

**Fig.1.** *Director fields around a large sphere, $R \gg K/W$, with perpendicular anchoring placed in a uniform nematic: (a) dipolar configuration with a hyperbolic point defect-hedgehog; (b) quadrupolar configuration with an equatorial disclination loop; (c) optical microscope texture of a sphere, $2R = 50\,\mu m$, immersed in a nematic E7 with large positive dielectric anisotropy $\Delta\varepsilon \approx 14$; (d) the same sphere under the action of an electric field of frequency 1.25 kHz and amplitude $E = 0.3\,V/\mu m$. When the electric field is periodically switched on and off (pulse duration 0.7 s, separation between pulses 3 s), the sphere moves to the right with an average velocity $0.2\,\mu m/s$; (e) equilibrium colloid with tangential surface anchoring; (f) fluctuative displacement of the colloid causes a restoring elastic force. Circular arrows in parts (a,b,e) stress axial symmetry of the system. In parts (c,d), photos and data by Israel Lazo.*

The geometry of distortions around large $R > \lambda$ particles depends on the shape of the particle and the orientation of the easy axis. A sphere with perpendicular anchoring placed into a uniform nematic creates a dipolar configuration, in which the matching between the local radial director $\hat{\mathbf{n}}_0 = \mathbf{r}/|\mathbf{r}|$ and the uniform far-field is achieved through a point defect, the so-called hyperbolic hedgehog [9], Fig.1a. The configuration is often described on the basis of the conservation law for topological charges. Note that since $\hat{\mathbf{n}} \equiv -\hat{\mathbf{n}}$, the charge of radial and hyperbolic hedgehogs can be chosen with an arbitrary sign, either +1 or -1. To facilitate proper calculation of charges in the conservation laws, it is convenient to introduce a vector field instead of $\hat{\mathbf{n}}$ [20]. Such a replacement is especially helpful in the description of colloids with complex geometries [21].

When the sample thickness is not much larger than $R$, or when there is an electric or magnetic field that aligns the director, the hyperbolic hedgehog spreads into a disclination ring [10, 22, 23], Fig. 1b. When the high-frequency AC electric field is periodically switched on and off, the hedgehog expands and shrinks because the dielectrically anisotropic LC tends to realign $\hat{\mathbf{n}}$ along the electric field. The periodic reconfiguration of the director results in the propulsion of the sphere in Fig.1c,d with the defectous side leading the way. The effect illustrates how anisotropy of the LC, in this case of dielectric permittivity, $\Delta \varepsilon = \varepsilon_\parallel - \varepsilon_\perp$, can be used to produce colloidal movement.

A new approach to control the director around colloids has been developed recently on the basis of photo-induced trans-cis isomerization of azo-dendrimers absorbed at the surface of glass spheres [24]; switching UV irradiation on and off causes the surface easy axis to realign and to trigger transformations between different director structures, such as dipolar, Fig.1a,c, quadrupolar Saturn, Fig.1b, and quadrupolar boojum pair, Fig.1e. When the spheres are replaced with elongated rod-like particles, UV-triggered rearrangement of the director causes reorientations of the rods and in some cases even their translation [25].

## 3. BROWNIAN MOTION OF COLLOIDS IN LIQUID CRYSTALS

In its simple realization, Brownian motion is observed as random displacements of a small particle in an isotropic fluid, controlled by the kinetic energy dissipation. The mean displacement is zero, but the average mean squared displacement (MSD) is finite, growing linearly with the time lag $t$, $\langle \Delta \mathbf{r}^2 \rangle = 6Dt$, where $D$ is the translational diffusion coefficient. For a sphere in a fluid of viscosity $\eta$, according to the Stokes-Einstein relation, $D = k_B T / 6\pi \eta R$.

In LCs, Brownian motion is anisotropic, with the coefficient $D_\parallel$ characterizing diffusion parallel to $\hat{\mathbf{n}}_0$ differing from the coefficient $D_\perp$ for perpendicular motion [26-29],

$$D_{\parallel,\perp} = k_B T / 6\pi \eta_{\parallel,\perp} R, \tag{1}$$

where the viscosities $\eta_\parallel \neq \eta_\perp$ depend on molecular orientation around the particle of the radius $R$. As the surface anchoring becomes stronger, the particle's diffusion slows down, as demonstrated by numerical simulations [30], due to the strengthening of the surrounding cloud of director distortions. Anisotropy of diffusion in LCs might be useful to mitigate the effect of complete randomization, known for isotropic dispersions of active colloids that are powered by local chemical reactions and do not rely on the external electric fields; however, the present author is not aware of any work on chemically powered colloids in LCs.

In addition to anisotropic character of Brownian motion, colloids in LCs exhibit also anomalous diffusion, $\langle \Delta \mathbf{r}^2 \rangle \propto t^\alpha$, where the exponent $\alpha$ is either smaller than 1 (subdiffusion) or larger than 1 (superdiffusion) [31]. These regimes are observed for motion both parallel and perpendicular to the overall $\hat{\mathbf{n}}_0$, when the time step between the measurements of the particle's positions is comparable to the relaxation time $\tau_{fl} \sim \eta R^2 / K$ of the director fluctuations around the particle; the subscript "fl" stands for "director fluctuations". At the time scales longer than $\tau_{fl}$, diffusion is normal, $\alpha = 1$, with mean-squared displacements parallel and perpendicular to the overall director following the linear dependencies $\langle \Delta x^2 \rangle = 2D_\| t$ and $\langle \Delta y^2 \rangle = 2D_\perp t$, respectively, with different diffusion coefficients $D_\|$ and $D_\perp$.

The time scales $\tau_{fl}$ separating normal anisotropic regime from the anomalous anisotropic behavior vary in a broad range. First of all, the experimental data [31] show that the values of $\tau_{fl}$ are different for diffusion parallel and perpendicular to $\hat{\mathbf{n}}_0$. Generally, $\tau_{fl}$ is determined by the viscoelastic properties of the medium and surface anchoring at the LC-particle interface. For standard thermotropic nematics such as pentylcyanobiphenyl (5CB) or cyanobiphenyl mixture E7, $\eta / K \sim 10^{10}$ s/m$^2$. Thus for $R = (1-5)$ μm, $\tau_{fl}$ is in the range 10 ms-0.3 s, but increases to 1-10 s for more viscous LCs, such as lyotropic chromonic LCs [31]. The anomalous diffusion is associated with the coupling of the director field $\hat{\mathbf{n}}(\mathbf{r},t)$ varying in space and time, to the velocity field $\mathbf{v}(\mathbf{r},t)$ of the LC; both $\hat{\mathbf{n}}(\mathbf{r},t)$ and $\mathbf{v}(\mathbf{r},t)$ are perturbed by the fluctuative motion of the particle. Weaker anchoring at the particle's surface is expected to mitigate the effect, since it would make the particle's impact on the director field $\hat{\mathbf{n}}(\mathbf{r},t)$ weaker.

To illustrate the mechanism of anomalous diffusion in a LC, consider a sphere with tangential anchoring, Fig.1e,f. A fluctuative displacement of the sphere, say, to its right by a distance $\delta \mathbf{x}$ temporarily increases the elastic energy of distortions on the right side and decreases it on the left side. The difference creates a restoring elastic force $\mathbf{F}_{sub} \propto -K \delta \mathbf{x}$ that pushes the particle backwards, in the direction opposite to $\delta \mathbf{x}$. Note that the characteristic times $\tau_{fl}$ of anomalous diffusion are much larger than the characteristic time $\sim \rho d^2 / \eta \sim 1$ μs of the isotropic hydrodynamic memory (time needed by the perturbed fluid of density $\rho$ and viscosity $\eta \sim 0.1$ Pa·s to flow over the distance $R$). Park et al demonstrated that both the anisotropic and anomalous features of the diffusion can be well mimicked by a stochastic process called fractional Brownian motion run with a multiscaling nonlinear clock [32].

One would expect further developments in the studies of translational and orientational diffusion of particles in LCs thanks to the recently proposed 3D image processing algorithm that can extract both position and orientation of the fluorescent particles from 3D confocal microscopy data [33].

Typical velocities of ACLCs discussed in this review are $v \sim 100\,\mu\text{m/s}$ or less; the particles are usually smaller than 100 $\mu$m. Therefore, the Reynolds number $\text{Re} = \rho v R / \eta$, which reflects the relative importance of the inertia and viscous terms in the dynamics, is very small. With the LC density $\rho \sim 10^3\,\text{kg/m}^3$ and viscosity $\eta \sim 0.1\,\text{kg m}^{-1}\,\text{s}^{-1}$ [7, 8], one estimates $\text{Re} \sim 10^{-5} - 10^{-4}$; the inertia can be neglected. Another important indicator is the Ericksen number $\text{Er} = \eta v R / K$ that measures the relative importance of the viscous and elastic forces. With $K = 10\,\text{pN}$, $\eta = 0.1\,\text{kg m}^{-1}\,\text{s}^{-1}$, $R = 50\,\mu\text{m}$, even a modest velocity $v = 2\,\mu\text{m/s}$ leads to $\text{Er} = 1$. For $\text{Er} < 1$, one can consider the Stokes drag on the sphere as a linear function of $v$. However, for Er>1, the coupling between the director and the velocity field makes the Stokes drag nonlinear, which might influence the dynamics of colloidal particles [34]. Numerical simulations by Stieger et al show a rather dramatic reconfiguration of topological defects around the colloids at Er>1 [35].

## 4. LINEAR AND NONLINEAR ELECTROPHORESIS IN LCs

The anisotropic character of LCs leads to numerous mechanisms by which a colloidal particle in a LC environment can be put into motion. These can be tentatively categorized as (i) transport by gradients of the order parameter in the absence of an external electric or magnetic field; (ii) transport by thermal expansion and thermal gradients; (iii) LC-enabled dielectro-phoresis, in which the gradients of the electric field are caused by director distortions of the LC; (iv) backflow-induced transport, associated with dielectric realignment of the director by a time-dependent electric field; an example is shown in Fig. 1c,d; (v) linear and nonlinear electrophoresis; (vi) transport through Quincke rotation; (vii) photoisomerization effects, etc. Most of these mechanisms were described in details in the recent review, including the dynamics caused by dielectric realignment [15*]. Below, we limit the consideration by the (v) electrophoresis and (vi) Quincke rotation effects, as they demonstrate interesting parallels and differences when compared to electrically driven colloids in isotropic media.

Electrophoresis is the motion of a charged particle in an electrolyte, powered by an externally applied uniform electric field. Both research and applications focus mainly on isotropic electrolytes such as water. The mechanism by which the electric field causes transport of particles in fluids depends primarily on how the electric charges are separated in space. In linear electrokinetics, the charges are separated through surface chemistry and form an electric double layer of thickness $\lambda_D = \frac{1}{e}\sqrt{\frac{\varepsilon\varepsilon_0 k_B T}{\sum_i c_i z_i^2}}$ (called the Debye screening length), typically on the order of 0.3-10 nm for aqueous electrolytes. Here $e$ is the electron charge, $\varepsilon$ is the dielectric constant of water, $\varepsilon_0 = 8.9 \, \text{pN/V}^2$ is the electric constant, $c_i$ and $z_i$ are the concentration and valency of ionic species $i$ in solution, respectively. Although the electrolyte in contact with a solid surface or particle remains an electrically neutral system, spatial separation of opposite charges in double layers is sufficient to cause electrokinetics, as illustrated in Fig. 2a for a charged colloid screened by counter-ions. An applied electric field induces a torque on the electric double layer, accelerating counter-ions in the fluid relative to the charges on the substrate until the motion is stabilized by the opposing viscous torque. The electrophoretic velocity is determined by the Helmholtz-Smoluchowski formula

$$\mathbf{v} = \frac{\varepsilon\varepsilon_0 \zeta}{\eta} \mathbf{E}, \qquad (2)$$

where $\zeta$ is the so-called zeta potential, i.e., electric potential at the slip surface near the solid surface; the electrolyte is assumed to be stagnant and of infinite viscosity in the narrow gap between the actual surface and the slip surface; at distances beyond the slip surface, the electrolyte's viscosity is assumed to adopt its bulk value $\eta$.

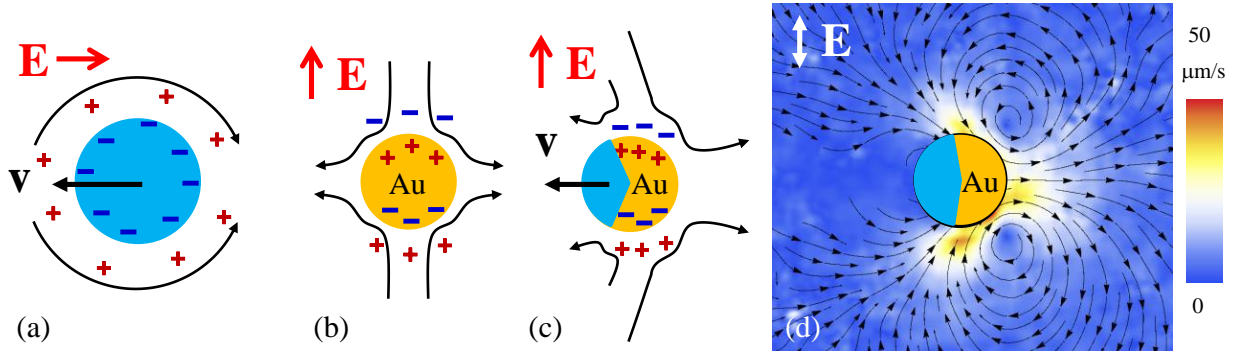

**Fig.2.** *Mechanisms of colloidal electrophoresis in isotropic electrolytes: (a) Linear electrophoresis of dielectric particle with surface charges screened by counterions; black curved arrows show the electro-osmotic flow around the sphere; if the sphere is free, it moves with velocity $v \propto E$. (b) electric field induces electric double layers around a metallic sphere and triggers four vortices of flow; as the pattern is symmetric with respect to the vertical and horizontal planes, the sphere does not move; (c) symmetry of the field-induced charges around a metallodielectric Janus sphere is broken, resulting in (d) asymmetric*

*electro-osmotic flows, as measured experimentally for a glass sphere with a hemispherical layer of gold, $2R = 50\,\mu m$, placed in a cell of thickness $60\,\mu m$; $E = 0.01\,V/\mu m$, see Ref.[50] for details..*

LCs always contain some amount of ionic impurities. They appear as residuals in chemical synthesis, through adsorption from adjacent media (such as alignment layers) and through injection of charges from electrodes in contact with the LC [36]. Electric conductivity of a typical LC with no added salts varies in a very broad range $10^{-7}$– $10^{-12}$ $\Omega^{-1}m^{-1}$ [36]. The Debye screening length $\lambda_D$ in LCs is about 0.1-1 $\mu m$, i.e. larger than the corresponding value in water. Thus in LCs, one should observe linear electrophoresis similar to the case of isotropic electrolytes; the only principal difference would be in the anisotropic character of electrophoretic mobility. The studies are limited to very few publications, the most recent are [28, 37-41]; the effect has been explored for possible applications in electrophoretic displays, as reviewed by Klein [42].

The linear relationship between velocities and the applied electric field implies that the flows of fluid and particles are possible only when driven by a DC electric field. An AC driving would produce no net displacement. DC driving does not allow one to produce steady flows persisting over extended period of time (longer than the time needed for ions from the electrolytic bulk to move towards the electrodes and to screen the applied electric field). Furthermore, since the velocity is proportional to the electric field in the linear effects, the corresponding electro-osmotic flows are irrotational, $\nabla \times \mathbf{u} = 0$, which makes it difficult to produce vortices needed, for example, in mixing. Among other problems encountered in linear electrokinetics, one often cites poor control of surface charges, relatively low velocities, and undesirable electrochemical reactions at the surfaces [43*, 44]. These considerations explain a growing interest to nonlinear phenomena, especially those in which the electrokinetic flows are proportional to $E^2$, thus allowing an AC driving.

As we will see below, nonlinear electrokinetics with quadratic field dependence is possible in both the isotropic and anisotropic LC electrolytes. In both cases, separation of electric charges of opposite polarity into different regions of space (i.e., formation of the "space charge") is induced by the applied electric field. If there is no field, the charges are not separated; this should be contrasted to linear electrophoresis, in which the charges are separated at the solid-fluid interface even in absence of the electric field. The field-induced density of space charge is in first approximation proportional to the electric field strength $E$. Therefore, the force acting on the space charge clouds, determined by the product of the charge and the electric field strength, should be proportional to $E^2$. It means that reversal of the field polarity does not change the polarity of the driving force (as the polarity of both the field and the field-induced charges changes to the opposite); hence one can use an AC driving to produce persistent flows of steady directionality. Since the viscous resistance is in first approximation independent on the applied electric field, the electrokinetic flows should feature velocities proportional to $E^2$. The advantages of the later behavior over the linear electrophoresis include (i) the possibility of using an AC field driving instead of the DC driving; (ii) achieving persistent AC-driven flows (as opposed to the flows in linear electrophoresis that are mitigated by ions moving under the DC field towards the electrodes and blocking them); (iii) generating vortices useful for applications such as mixing at small scales. Although electrokinetics with the quadratic velocity-vs.-field behavior is possible in both the isotropic and LC electrolytes, the underlying mechanisms of space separation are different and can be considered as opposite/complementary to each other. In the isotropic case, separation of charges is promoted through the properties of the solid part, say, the colloidal particle or the wall of the microfluidic chamber. In the LC case, the space charge is created because of the properties of electrolyte, namely, spatially varying director field. We first consider an example of charge-induced electrophoresis of a particle in an isotropic electrolyte. The electric double layers in this case are created by the applied electric field around a highly polarizable (metal) particle; the condition of propulsion is an asymmetry of the colloidal particle, either in shape or in physical properties.

Consider a metallic sphere placed in an isotropic electrolyte, acted upon by a uniform electric field **E** [45, 46]. The field polarizes the sphere. The field also drives the ions through the electrolyte towards the sphere, inducing a double layer, Fig.2b. The double layer expels the field lines, creating a component of the field that is tangential to the sphere's surface. This component drives the induced charges, thus producing an electro-osmotic flow around the particle. If the particle is homogeneous, the flows are of quadrupolar symmetry, with four vortices, Fig.2b. Although such a sphere produces flows, it cannot move itself. Propulsion becomes possible when the particle is asymmetric, for example, a metallodielectric Janus sphere. In this case, the fore-aft symmetry is broken, as the field-induced charge and the electro-osmotic flows near the metallic part are stronger than near the dielectric part, Fig.2c,d. Asymmetry of the electro-osmotic flows leads to propulsion of the Janus particle [43*], as discovered experimentally by Velev et al [47]. The effect is observed not only for the Janus spheres [48] but also for doublets [49]. The broken symmetry of electro-osmotic flows around metallodielectric spheres has been confirmed in recent experiments, Fig.2d [50, 51]. For highly polarizable metallic particles, the field-induced density of space charge is proportional to the applied field, thus the ICEP velocities grow with the square of the field, allowing one to use an AC driving [45, 46] [47*].

Electrophoretic motion in isotropic electrolytes imposes certain requirements on the particle that must be either charged or highly polarizable; the surrounding isotropic electrolyte plays a supportive role, supplying the counterions. When the isotropic electrolyte is replaced with a LC, these "isotropic" mechanisms are still functional. However, the anisotropy of LCs brings about a new mechanism of charge separation, electrophoresis [38*], and electro-osmotic flows [52**] called liquid crystal-enabled electrokinetics (LCEK). We remind that the term "electrokinetics" combines the concepts of electrophoresis (usually defined as motion of a solid particle with respect to the fluid powered by a uniform electric field) and electro-osmosis (motion of the fluid electrolyte with respect to the solid surface). The mechanism is rooted in the properties of the anisotropic LC electrolyte [38*, 52**], and lifts many limitations on the properties of the particles, as discussed below.

Consider a sphere with perpendicular anchoring in a uniform LC, placed between two plates with tangential anchoring. Suppose the cell is shallow, so that in response to the surface anchoring conditions, the LC creates a disclination ring around the sphere that matches the local radial director with the uniform far-field, $\hat{\mathbf{n}}_0 = \{1,0,0\}$, Fig.3a. Suppose the electric field $\mathbf{E} = (E_x, 0, 0)$ is applied along the $x$-axis and that the dielectric anisotropy is zero, so that the electric field does not realign the director; its only action is to drive the ions. If $\Delta\sigma > 0$, the ions prefer to move along the director lines. In Fig.3a, for the shown polarity $E_x > 0$, positive ions accumulate on the left side of the sphere, while the negative ones settle on the right side. The density of the induced space charge should be proportional to the electric field, $\rho \propto E_x$.

Once the charges are separated in space, the electric field drives them by imposing a Coulomb force of density $f \propto \rho E_x$, which yields an electroosmotic flow of the nematic around the sphere, Fig.3a. Reversing the field polarity alters the sign of the induced charge $\rho \propto E_x$, as clear from Fig.3a, but the product $f \propto \rho E_x \propto E_x^2$ remains polarity-insensitive. It means that the forces and flows are also polarity independent, growing as $E_x^2$. The experimentally determined flow map of flows around a sphere in a LC, driven by an AC electric field, Fig.3b, confirms the expected outcome. The flow patterns and velocities are determined by adding small fluorescent polymer tracers to the LC and characterizing their trajectories under an optical microscope.

Different types of surface anchoring produce different polarities of space charges. For example, a tangentially anchored sphere, Fig.3c, acted upon by the horizontal electric field $E_x > 0$, acquires a negative charge on the left side and a positive charge on the right side, Fig.3c. The polarity is opposite to the one around the perpendicularly anchored sphere in Fig.3a. In other words, the induced space charge is sensitive to the sign of the director gradients $\partial\varphi/\partial y$, where $\varphi$ is the angle between the unperturbed $\hat{\mathbf{n}}_0$ and the actual $\hat{\mathbf{n}}$, Fig.3a; the y-axis is directed along the vertical direction in the plane of Fig.3, being perpendicular to the far-field $\hat{\mathbf{n}}_0 = \{1,0,0\}$. As a result, the two spheres produce opposite polarities of flows. The electro-osmotic flow is of a "puller" type around the normally anchored sphere, i.e., the inward velocities are collinear with $\mathbf{E}$, Fig.3b, and of a "pusher" type around the tangential sphere, Fig.3d. Experiments [52**] for both types of anchoring at the surface of glass spheres in shallow LC cells confirm the prediction of "puller" and "pusher" flows, Fig.3b and Fig.3d, respectively. In both cases the flows are of quadrupolar symmetry and thus the spheres do not move. However, the quadrupolar symmetry of the director can be easily broken, for example, when the perpendicularly anchored sphere is placed in a relatively thick nematic cell, Fig.1a. The dipolar director pattern translates into dipolar symmetry of the induced charges and flows. If the dipolar particle is immobilized, the induced electro-osmotic flows around it pump the LC; for the shown patterns of flows in Fig.3e, pumping is from the right side of the chamber to the left, Fig.3f; depending on the concrete type of the LC, frequency of the field, etc., opposite polarity of pumping can also be observed. If the particle in Fig.3e is not attached to the substrate, it moves electrophoretically, from left to right, with a velocity proportional to $E_x^2$, as discussed below.

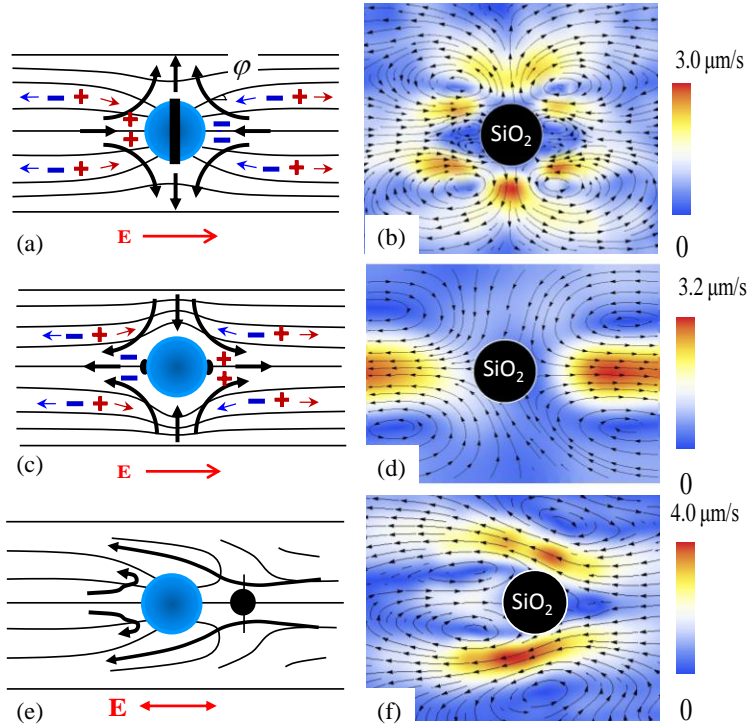

*Fig.3. Anisotropy of conductivity separates charges and creates electro-osmotic flows in a LC distorted by presence of (a,b) a sphere with perpendicular anchoring and quadrupolar Saturn ring director configurations; (c,d) sphere with tangential anchoring and quadrupolar director distortions; (e,f) sphere with perpendicular anchoring and dipolar director distortions, attached to the bottom of the cell by glue; asymmetry of electro-osmotic flows produces pumping of the LC*

*from right to left. Left column shows electro-osmotic flows of steady directionality (thick arrows) at the background of director fields (thin lines). Right column shows experimentally determined electro-osmotic flows around immobilized glass spheres; $2R = 50\,\mu m$; $E = 0.026\,V/\mu m$, $f = 5\,Hz$. Note the DC character of the electric field in parts (a,c) and AC in parts (b,d,e,f). Modified from Ref. [52\*\*].*

The charge density induced by the electric field in a distorted LC can be calculated under the approximation of weak anisotropies of conductivity and permittivity ($\Delta\varepsilon \ll \bar{\varepsilon}$, where $\bar{\varepsilon} = (\varepsilon_\parallel + \varepsilon_\perp)/2$, $\Delta\sigma \ll \bar{\sigma}$, $\bar{\sigma} = (\sigma_\parallel + \sigma_\perp)/2$) and weak director gradients as [52\*\*]:

$$\rho(x,y) = \varepsilon_0 \bar{\varepsilon} \left( -\frac{\Delta\sigma}{\bar{\sigma}} + \frac{\Delta\varepsilon}{\bar{\varepsilon}} \right) \frac{\partial\varphi}{\partial y} E_x. \qquad (3)$$

The charge density is proportional to the applied field $E_x$ itself, but also to the properties of the LC, namely, (i) the director gradients $\partial\varphi/\partial y$, (ii) anisotropies $\Delta\sigma$ and $\Delta\varepsilon$. Thus liquid crystal-enabled electrokinetics, or LCEK, [38\*, 52\*\*], in which the space charge occurs because of the properties of the medium, is very different from the induced-charge electrokinetics in isotropic electrolytes, where it is the properties of the particle (or a solid wall) that are responsible for the separation of charges.

The amplitude of electro-osmotic velocities around spheres in Fig.3a-d, or, equivalently, the electrophoretic velocity of the free dipolar sphere in Fig.3e, follows from the balance of the driving Coulomb force $f \propto \rho E_x$ and viscous resistance $\eta v / R^2$ [52\*\*]:

$$|v| = \frac{\alpha \varepsilon_0 \bar{\varepsilon} R}{\eta} \left| \frac{\Delta\sigma}{\bar{\sigma}} - \frac{\Delta\varepsilon}{\bar{\varepsilon}} \right| E^2. \qquad (4)$$

The dimensionless coefficient $\alpha \sim 1$ is introduced to account for the details of director configuration, anchoring strength and direction, replacement of anisotropic LC viscosity with its average value $\eta$ and for other approximations, such as using $1/R$ as a measure of director gradients. Experiments show that the LC enabled electro-osmotic velocities are indeed proportional to the radius of the particle $R$ and to $E^2$ [52\*\*]. The latter allows one to produce steady flows by AC driving. The effect is sensitive to the frequency of the applied electric field, as the characteristic time of charge build up is finite; experimentally, the maximum of electrokinetic velocities, on the order of $10\,\mu m/s$, is achieved for frequencies in the 5-500 Hz range. At higher frequencies, the conductivity effects are less effective and one can use dielectric driving, as in Fig.1 c,d.

The flows shown in Fig.3 are created with dielectric (glass) particles; there are no metal/conductive elements in the LC cell, except for the two electrodes. It has some important implications for potential applications. For example, as demonstrated by Hernàndez-Navarro et al [53**] and by Sasaki et al [54**], the LC-enabled electrophoresis can be used to transport droplets of fluids that are not miscible with the LCs, such as water with surfactant additives [53**], silicon oil [54**], etc., provided the surface anchoring is supporting the dipolar configuration around the droplet, Fig.1a. This possibility opens the door for various type of "microreactors" settings. Furthermore, the particle's trajectory can be controlled through the pre-designed director configuration in the cell, as demonstrated for circular orbiting of colloids placed in a cell with director patterns [38*]. The overall director alignment in the cell can also be switched, as demonstrated by Hernàndez-Navarro et al [55, 56], who used trans-cis isomerization of azobenzene derivatives to drive the director realignment. The trajectories of the electrophoretically active colloidal swarms can thus be controlled by light irradiation [55, 56].

The mechanism of electrokinetics described above does not require the director distortions to be induced exclusively by the colloidal particles. One can imagine, for example, that the patterns similar to those shown in Figs. 1a,c,e are imposed on the director through a patterned surface alignment. The charge separation in the presence of the electric field would follow the same qualitative pattern. If colloidal particles with an arbitrary surface anchoring are dispersed into a LC, they would be carried along by these flows; to move, these particles do not need to create a dipolar director field around themselves. This patterned LCEK has been demonstrated recently by Peng et al [57*]. The director distortions were created through photo-patterning of surface anchoring at the bounding plates of the cell. The predesigned patterns allowed an applied electric field to separate charges in space, similarly to the case of particle-induced distortions shown in Fig.3. The ensuing electrokinetic flows were demonstrated to carry practically any cargo (solid, fluid or even gaseous) placed in the chamber with the director patterns, regardless of the surface anchoring, polarizability, electric charge, etc. [57*].

In the consideration of electrokinetics in Fig.3, the electric field did not change the director orientation. Generally, a LC does change its orientation in the applied electric field, thanks to (i) anisotropy of dielectric permittivity $\Delta\varepsilon$ (the so-called Frederiks effect), (ii) flexoelectricity, and (iii) anisotropy of conductivity. The dielectric reorientation, illustrated in Fig.1c,d for a nematic E7 with a relatively large positive dielectric anisotropy $\Delta\varepsilon=13.8$ can be suppressed, if needed, by simply using materials with a small $\Delta\varepsilon$ [52**]. The flexoelectric polarization exists even when the field is absent, and one would expect it to produce corrections to the electrophoretic velocities that scale as $E$ or $E^3$ rather than $E^2$. Finally, the conductivity anisotropy can destabilize the uniform director in the cell even in the absence of the colloidal particles, when the field exceeds some threshold value. Sasaki et al [54**] explored LC-enabled electrophoresis in a uniformly aligned planar cell filled with the nematic of negative dielectric anisotropy $\Delta\varepsilon = -0.65$. The field was perpendicular to the overall director. At relatively low applied voltages, the dipolar particle, Fig.3e, exhibits a steady linear motion with velocity $v \propto E^2$; the overall director in the cell remains uniform. However, once the field exceeds some threshold value, an electrohydrodynamic instability develops and the particle's trajectory becomes sinusoidal in the vertical plane, being shaped by both the electrohydrodynamic convection and the LC-enabled electrophoresis; the particle's velocity considerably increases. In the same work, Sasaki et al [54**] demonstrated an interesting collective effect. Because of the long-range elastic interactions, the dipolar particles form elongated chains parallel to the overall director. When such a chain, comprised of 30 glass microspheres, is subject to the perpendicular electric field above its threshold value, it develops a sinusoidal caterpillar-like shape [58] and moves along the wave-vector of periodic electrohydrodynamic rolls; the velocity increases with the field. The colloidal caterpillars can transport a "cargo", i.e. a particle that is by itself electrophoretically inactive, such as a symmetric rod.

## 5. QUINCKE ROTATORS AND COLLECTIVE TRANSLATIONS

In the limit of DC driving, a good approach to drive colloidal particles through both the isotropic and the LC phases is to use the Quincke effect [59, 60*]. The Quincke effect is defined as a rotation of a dielectric particle neutrally buoyant in an isotropic fluid under the action of a DC field [61]. The mechanism is related to a difference in the characteristic times of electric charge relaxation within the particle and in the surrounding fluid. If the first is longer than the second, the effective electric dipole is anti-parallel to the applied DC field, Fig.4a. Such an orientation is unstable and, when the field is sufficiently high to overcome the viscous friction, the particle starts to spin. If its location is not exactly in the middle plane of the cell, interaction with the substrates converts rotation into translation, similarly to the motion of a spinning coin rolling on its edge along a solid substrate, Fig.4a [60*]. The translational velocity of Quincke rotators is proportional to the angular velocity $\Omega$ of spinning; both are growing with the applied field as $v \propto \Omega \propto \sqrt{E^2 / E_{th}^2 - 1}$, where $E_{th}$ is the threshold field [60*].

An interesting collective effect is observed in smectic A cells, Fig.4b-d. The smectic A exhibits the same orientational order as the nematic does, with an additional one-dimensional positional order along $\hat{\mathbf{n}}$

; the molecules form 2D fluid layers stacked on top of each other. In the described experiments [60*], **n̂** is perpendicular to the glass plates bounding the LC, thus the rotating Quincke particles move in the direction perpendicular to **n̂** in an effectively isotropic 2D environment. If the smectic A cell contains air bubbles, the Quincke rotators are attracted to the smectic A meniscus of a circular shape and start orbiting the air inclusions. If the number of particles is low, Fig.4b, they move in either direction, clockwise and counterclockwise, colliding and reversing the direction. However, as the number of trapped particles increases, they settle into a collective macroscopic unidirectional orbiting, either clockwise or counterclockwise, Fig.4c,d. The presence of air bubbles with circular "tracks" for the Quincke rotators facilitates a transition from random translations to an ordered macroscopic motion, similar to the predictions of the Vicsek model [62]. Recently, the effect of collective motion was demonstrated in a similar experimental system, with Quincke rotators moving in an isotropic fluid along an elongated track; the motion becomes ordered when the concentration of particles exceeds a certain threshold [63]. Further experiments with LC environment would be of interest, as the LC can help to tune the interactions among the self-propelled Quincke rotators.

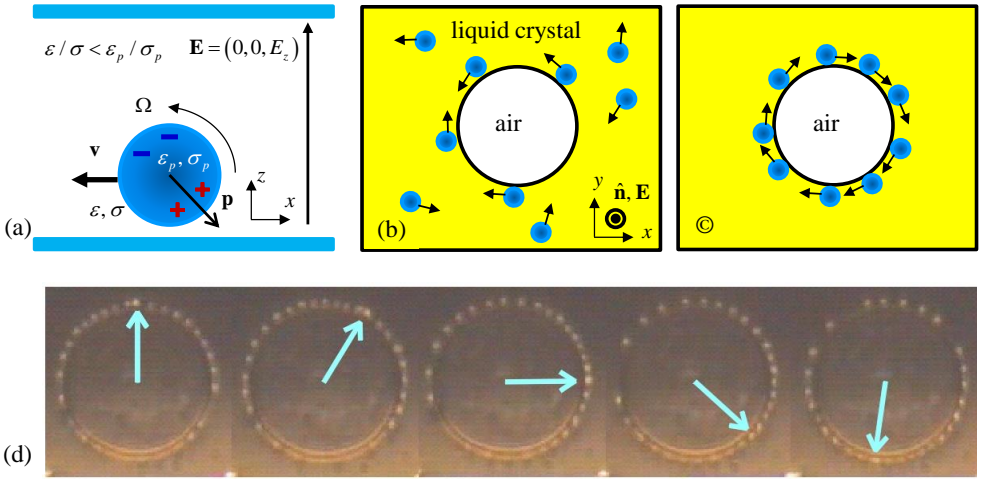

*Fig.4. (a) Dielectric particle in a fluid medium develops rotation if the charge relaxation time in the particle ($\propto \varepsilon_p / \sigma_p$) is larger than the charge relaxation time of the fluid ($\propto \varepsilon / \sigma$). If the Quincke rotator is located closer to one substrate than to the other, rotations cause translational motion; (b) Quincke rotators attracted to the circular liquid crystal-air bubble interface and move clockwise and counterclockwise if their concentration at the interface is low; (c) at high concentrations, facilitated by attraction of particles to the meniscus in the smectic A phase, the Quincke rotators settle into a macroscopically coherent orbiting; (d) optical microscopy*

*texture of collective clockwise orbiting of an air bubble (diameter $90\,\mu m$) by the Quincke rotators of diameter $2R = 4.5\,\mu m$ in the smectic A. In parts (b,c,d), both the director and driving DC electric field are perpendicular to the field of view. Snapshots in (d) are taken every 1/3 s. The arrows point towards the same orbiting glass sphere. For more details and motion pictures see Ref. [60*].*

To conclude this section, note that the direction of motion of colloidal particles in the LC environment depends on both the direction of the electric field and the local orientation of the director. Depending on the mechanism of propulsions, the trajectories might feature different preferences. Trajectories of Quincke rotators in a smectic A slab are perpendicular to both the overall director and the applied electric field [60*]. In the case of LCEK in uniform nematic cells, colloids move parallel to the overall director and either parallel (in a LC with positive dielectric anisotropy [38*]) or perpendicular (negative dielectric anisotropy [28]) to the electric field. However, trajectories perpendicular to the director are also possible, for example, if one uses linear electrophoresis in a LC environment [28]. In the nematic cells with pre-patterned director distortions, the LCEK flows can be either parallel or perpendicular to the local director [57*]. These differences underline a rich spectrum of design opportunities for transport of colloids in LCs.

## 6. BACTERIA IN LYOTROPIC CHROMONIC LIQUID CRYSTALS

Active matter formed by self-propelled particles exhibits fascinating dynamic effects and patterns, shaped by interactions (often short-ranged) of the active particles among themselves and with the environment. Most studies deal with self-propelled particles placed in a Newtonian isotropic fluid, so that the interactions are of the hydrodynamic and excluded volume types. One of the fascinating effects in these isotropic systems is an emergence of orientational order due to activity. In the celebrated Vicsek model [3, 62], the order parameter is defined through the average velocity vector of the system of self-propelled particles. When the concentration of active particles increases, the system undergoes a phase-transition-like change from randomly moving particles to particles moving along the same macroscopic direction.

LCs, used as an environment for active colloids, allows one to explore a different situation, when the environment is anisotropic and imposes a "sense of direction" for any concentration of active particles and for any level of their activity, including zero. If the particles do not move, they can still align with respect to the surrounding LC director, provided their shape is not an ideal sphere [64]. Even when the particles are spheres, they show long-range anisotropic interactions mediated by the LC elasticity and form chains that are parallel [9] or tilted [65] with respect to the director.

The orientational order of the LC medium and the activity of self-propelled particles in it can be tuned independently. Such a tunable system can be even transformed into two reference states, namely, (i) an isotropic active state (in which the orientational order of the LC is melted but the activity is preserved) and (ii) inactive orientationally ordered equilibrium LC state (by switching off the activity of particles). Quincke rotators in the isotropic, nematic, and smectic A phases discussed above are one realisation of such a system. In this section, we consider a different example, in which the particles placed in the LC are self-propelling bacteria. The LC modifies the behavior of the bacteria, while the activity of bacteria modifies the orientational order of the LCs. Depending on the concentration of bacteria, we distinguish individual particle effects (low concentration of bacteria) and collective phenomena (high concentration). In the latter case, the system has been called "living liquid crystals" [66**]; it will be considered in the next section.

The LC in question is the so-called chromonic lyotropic liquid crystal (LCLC) that represents a water dispersion of plank-like organic molecules with a relatively flat rigid polyaromatic core and polar groups at the periphery; for recent reviews, see [67-69]. In water, these molecules aggregate face-to-face, stacking on top of each other, in order to minimize the areas of unfavorable contact with water. The molecules form rod-like aggregates of the so-called H-type (molecular planes are perpendicular to the axis of the aggregate) with the polar groups at the water interface. The stacking distance is about 0.33-0.34 nm. The stacking principle, density of polar groups, and typical geometrical parameters (the cross-section of the aggregate can contain one, two, sometimes more individual molecules) make the chromonic aggregates similar to double-strand B-DNA molecules. The important difference is that in LCLCs, there are no chemical bonds between individual molecules to fix the length of aggregates; the latter is determined by the balance of entropy and the "sticking" energy of attractive interactions of the molecular cores. Unlike their surfactant-based micellar and thermotropic counterparts, the LCLCs are not toxic to biological cells [70].

The most widely used LCLC compound is disodium chromoglycate (DSCG) that forms a nematic phase at room temperatures for concentrations in the range 0.3-0.45 mol/kg [68], or between about 12.5 and 18 wt%. Water solutions of orientationally ordered DNA molecules can also be used as an LC environment for bacteria [71]. Upon heating, DSCG dispersion experiences a phase transition from the nematic to an isotropic state, first forming a broad biphasic region (of width 4-9 °C, depending on concentration) of the coexisting isotropic and nematic phases [72], then a homogeneous isotropic phase.

When placed in the nematic phase of DSCG, elongated flagellated bacteria, such as *Escherichia coli* (*E. coli*) [73], *Proteus Mirabilis* (*P. Mirabilis*) [74*,75**,76], and *Bacillus subtilis* (*B. subtilis*) [66**, 77] all preserve their ability to swim, although to a different degree, see the discussions in [76] and [77]. All three types of bacteria are of a similar geometry. For example, *B. subtilis* studied by Zhou et al [66**, 77] have a rod-like "head" of length $L=(4-7)$ μm and diameter 0.7-1 μm, to which six or more flagella are attached, each of length 6-20 μm and cross-sectional diameter 20 nm. A flagellum is of stiff left-handed helicoidal shape. Hydrodynamic attractions among flagella produce a single polar rotating bundle posterior to the swimming organism [78]. When the bacterium swims in the birefringent LC, this bundle becomes clearly visible under a polarizing optical microscope with an attached LC PolScope unit (which allows one to map the local optical retardance), as a periodic wave of variable retardance with a pitch $\tilde{\lambda}=$ 2 μm, Fig.5a. *B. mirabilis* in the studies by Mushenheim et al [74*, 75**, 76] are somewhat shorter, with the head about 3 μm long. Typical velocities in the nematic phase of DSCG range from ~3 μm/s for *E. Coli* [73], to 8 μm/s for *B. mirabilis* [76] and 8-14 μm/s for *B. Subtilis* [66**, 77].

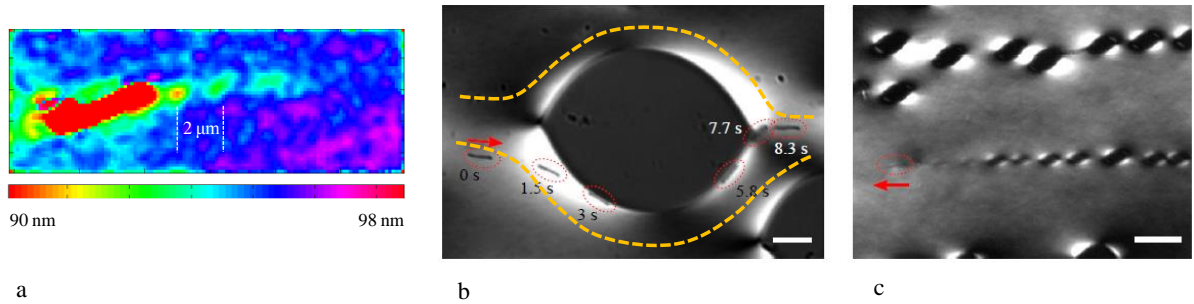

**Fig. 5.** *Individual bacteria B. subtilis in a nematic LCLC formed by DSCG dispersion in water. (a) polarizing microscopy image of the bacterium with a head on the left and bundle of flagellae on the right; the wavy character of the bundle with a period 2 μm is visualized by nematic birefringence in the LC PolScope map of optical retardance; (b) bacterium (enclosed by a red ellipse) moving from left to right around an isotropic tactoid along the curved director (schematically shown by the orange dashed lines); scale bar 10 μm (c) a bacterium moving from right to left in the nematic close to the melting point, creates a chain of isotropic tactoids in its wake; scale bar 20 μm. Reproduced from Ref. [66**].*

If the LC is aligned in the planar fashion, with the director pointing along a single direction $\hat{\mathbf{n}}_0$ in the plane of the cell, the bacteria move along $\hat{\mathbf{n}}_0$ [66**, 71, 73, 74*, 75**, 76, 77]. The main reason is the surface anchoring of the bacteria: when the bacteria are immobilized, their rod-like bodies are still parallel to $\hat{\mathbf{n}}_0$, suggesting tangential easy axis [66**, 71, 76]. Tumbling is strongly suppressed by the nematic elasticity, as realignment of the bacterium perpendicularly to the overall director requires overcoming a barrier on the order of $KL \sim 5\times 10^{-17}$ J (the Frank constants of DSCG solutions is on the order of 10 pN [79, 80]), much higher than the typical torque $10^{-18}$ J of the flagella motor that powers tumbling in an isotropic medium.

When the director is distorted, the bacteria's trajectory is no longer linear. The director distortions can be created by heating the DSCG solution into the biphasic region, in which one of the two phases exists as islands, called tactoids, surrounded by the "sea" of the other phase. At lower temperatures, isotropic tactoids shaped as an American football distort the nematic director in the surrounding matrix [72], Fig.5b. Zhou et al [66**] reported that *B. subtilis*, swimming in the nematic phase, approaches the nematic-isotropic interface but does not necessarily cross it; instead it moves along the curvilinear director lines at the interface and then escapes the tactoid at its "cusp", Fig.5b. Mushenheim et al [74*], working with *P. mirabilis*, which are about two times shorter than *B. subtilis*, found that, in addition to the scenario above, individual *P. mirabilis* very often cannot escape the strongly deformed cusp region and have to wait until more of their partners accumulate there to gather a critical mass (typically 2-10 bacteria) sufficient for a successful group escape. When the biphasic region is kept at somewhat higher temperatures, the nematic tactoids are surrounded by the isotropic melt. The bacteria from within the nematic tactoids are directed towards the two cusps region where the director experiences strong splay; from there, they escape into the isotropic melt [74*]. Preferential partitioning of *P. mirabilis* in the isotropic regions was demonstrated to be sufficiently strong to allow one to concentrate them in the isotropic tactoids that are shrunk by decreasing the temperature [74*].

The accumulation of particles at defectous regions either in the bulk or at the surface of a LC is known to be a result of the balance of elastic and anchoring forces (a colloidal particle is usually trapped at the defectous region as it replaces the deformed region of a LC with itself, when the balance of changing elastic and surface anchoring brings the total energy of the system down [81, 82]). The observation of the individual and group escapes poses a new interesting question of how active particles interact with the spatially varying environment. Note that the escape of bacteria from the cusp region might be controlled not only by the director configuration but also by the diffuse character of the isotropic-nematic interface, in which the scalar order parameter (measuring the degree of orientational order) changes over the length scale on the order of 10 μm [72].

Nematic elasticity modifies the behavior of motile bacteria not only in the sense of helping them to select the direction of motion but also by promoting aggregation. *B. mirabilis* often dynamically assemble end-to-end into moving chains, either in the bulk [76] or at the nematic-isotropic interface [74*]. The separation between two bacteria is rather small, about 0.3 μm. Immobilized versions of *B. mirabilis* aggregate irreversibly, but the motile chains can dissociate back into monomers [76]. The *B. mirabilis* chains in the nematic bulk are slightly tilted with respect to the far-field director [76], as expected for the equilibrium chain of elastically interacting elongated (ellipsoidal) particles with tangential surface anchoring [83].

*B. subtilis* in similar planar cells of DSCG interact differently from *B. mirabilis*. Instead of the end-to-end aggregation, they exhibit transient parallel alignment side-by-side with synchronization of their velocities [77]. The synchronization is likely caused by the dynamic elastic director pattern. The bacterium in motion creates a complex wave-like distortion, Fig.5a, that is closer to a dipolar rather than quadrupolar configuration. For example, near the temperature of the nematic-isotropic transition, the shear stress produced by the flagella is so strong that the nematic experiences a local transition into an isotropic state and leaves isotropic tactoids in the wake of a moving *B. subtilis* [66**], Fig.5c In the general classification of colloid-induced distortions in nematics, proposed by Pergamenshchik and Uzunova [84, 85], the closest type is the so-called chiral z-dipole. The z-component is associated with the bend and the chiral part with both twist and bend. Two parallel bodies of a characteristic width $\omega$, with the shape of chiral z-dipoles, placed side-by-side in a nematic LC at a distance $a$ interact through the elastic potential $U = 12\pi K \tilde{\lambda}^2 \omega^2 \left( \tilde{C}^2 - \tilde{\gamma}^2 \right) / a^3$, where $\tilde{\lambda}\omega\tilde{C}$ is the chiral strength and $\tilde{\lambda}\omega\tilde{\gamma}$ is the strength of the z-dipole; $\tilde{C}$ and $\tilde{\gamma}$ are shape-dependent numbers, of the order of 0.1-10, and $\tilde{\lambda}$ is the chiral pitch. For $\tilde{\lambda} = $ 2 μm, $\omega = $1 μm, $a=$2 μm, the estimate is $U = 2\times 10^{-17} \left( \tilde{C}^2 - \tilde{\gamma}^2 \right)$ J, potentially many orders of magnitude higher than the thermal energy $4\times 10^{-21}$ J and comparable to or larger than the torque of the flagellar motor ($10^{-18}$ J). The interaction can be either attractive or repulsive, depending on the sign of $\left( \tilde{C}^2 - \tilde{\gamma}^2 \right)$, thus bacteria of even a somewhat similar shape might show very different LC-mediated interactions.

The velocity of the bacteria in the LC is not much smaller than the velocity in water, by a factor of about 1.5, 14 μm/s vs 21 μm/s [77]. This relatively small difference comes as a surprise since the twist and splay viscosities of the LC are about three orders of magnitude higher than the viscosity of water [66]. The standard model of bacterial swimming in an isotropic Newtonian fluid is based on the difference in the effective drag coefficients $\xi_{//}$ and $\xi_{\perp}$ for the parallel and perpendicular displacements of the flagellar bundle [86]. In the LC, the difference in drag coefficients for the parallel and perpendicular displacements is assured by the anisotropic character of the medium. For diffusive motion, the ratio $\xi_{//}^{LC} / \xi_{\perp}^{LC}$ can change in a broad range, from 1.1 to 2.5 [29, 31]. Therefore, the anisotropy of the LC might help the bacteria to swim [77]. Note that the comparison of behavior of bacteria in a LC and water was performed in Ref. [77].

when both systems contained nutrients (the so-called Terrific Broth solution); the latter does not change the nematic state of the DSCG at room temperatures.

The rotating bundle of flagella causes dynamic distortions of the LC, Fig.5a, that can be used to measure its frequency, through direct observations of the LC birefringent pattern under a polarizing microscope; the bundle rotates with frequency in a range between 6 and 20 Hz, while the head rotates with frequency around 7 times smaller in the opposite direction. The translational velocity of the bacteria shows a linear dependence on the frequency $f$ of flagella rotation, expressed in Hz units, $v = f \times 0.8 \,\mu\text{m}$; the phase velocity $v_f$ of the flagella wave is about two times higher than $v$. In contrast, in the isotropic fluid, $v_f / v \approx 10$. The difference suggests that the fluid flow created by a bacterium in the LC does not spread much around the bacterial body, being instead localized close to the bacterial trajectory, within the distance estimated roughly as a few micrometers; reconstructed flow patterns support this conclusion [77]. The concentrated character of flow is also confirmed by the effect of cargo transportation. When there is a small colloidal particle in front of *B. subtilis*, the bacterium pushes this particle forward over a long distance, sometimes by hundreds of micrometers [77]. The particles start to move when the distance to the bacterium is as large as 50-80 $\mu\text{m}$. If the particle is off the bacterial trajectory by only a few micrometers, the bacterium swims over without picking it up. *B. mirabilis* also shows an ability to transport cargo along the unidirectional path or even along a complex patterned director field [87].

The discussion above referred to a situation when the LC alignment in a sandwich type cell is planar. When the alignment is homeotropic ($\hat{\mathbf{n}}_0$ is perpendicular to the plates), new effects are observed for *B. mirabilis* [75**]. First, if the LC cell is a few times thicker than 3 $\mu\text{m}$ (the normal length of *B. mirabilis*), the bacteria gets trapped at the bounding plates, being perpendicular to them. The result implies that the elastic torque imposed by the director is stronger than the hydrodynamic torque that makes the bacteria swim parallel to flat bounding plates in isotropic solutions. The strength of the elastic torque can rectify the polarity of bacterial motion. In a clever experiment [75**] demonstrating rectification, two opposite plates of the nematic cell were treated for antagonistic surface alignment, perpendicular at one plate and planar at the other plate. This director field, resembling a rounded capital letter L, implies a polar character of the director projection $\hat{\mathbf{n}}_{xy} \neq -\hat{\mathbf{n}}_{xy}$ onto the plane of the cell. As a result, bacteria preferentially swim from the top to the bottom along the "L" – shaped director lines, but not in the opposite direction. In experiments with anomalously long *B. mirabilis*, $L = (10 - 60) \,\mu\text{m}$, placed in the homeotropic cells of thickness 10 $\mu\text{m}$, the bacteria swim in the direction perpendicular to $\hat{\mathbf{n}}_0$, with the velocity about 1.5 times smaller than the velocity parallel to $\hat{\mathbf{n}}_0$ in the planar cells [75**], a manifestation of anisotropic viscous drag in LCs.

The theoretical treatment of bacterial swimming in a LC is of enormous difficulty. The equilibrium static properties of the LC can be described in terms of Frank elasticity, surface anchoring, and anisotropic coupling to the external electric fields. For the description of dynamic phenomena, one needs to add anisotropic viscous response, which might lead to the so-called flow-aligning and tumbling regimes. The relative role of elastic, anchoring, and viscous contributions depends on the characteristic frequencies and length scales in the system. The first theory of microscale locomotion in a LC has been recently proposed by Krieger, Spagnolie, and Powers [88*]. The theory considers the so-called Taylor's swimming sheet, a one – dimensional swimmer with a small-amplitude wave of either transverse or longitudinal type, placed in a 2D hexatic LC. The hexatic LC captures the essential properties of a nematic, such as Frank elasticity, rotational viscosity and surface anchoring. The results clearly show that the swimming efficiency and speed depend on the Ericksen number $\mathrm{Er} = \eta v R / K$ and the de Gennes-Kleman anchoring length $K/W$. For example, for transverse waves with weak anchoring, the swimming velocity decreases as $\mathrm{Er}$ becomes higher. This dependency of velocity on $\mathrm{Er}$, however, is mitigated by a stronger surface anchoring.

## 7. LIVING LIQUID CRYSTALS

The results discussed above refer to individual bacteria or small clusters of them. Below, we discuss a collective behavior that emerges at higher concentrations of bacteria in LC, on the order of $10^{15}$ bacteria/m$^3$, i.e., 1 bacterium in a cube with a side length about 10 µm.

*B. subtilis* is an aerobic bacterium, the motility of which is controlled by the amount of dissolved oxygen [89]. Bacteria stop swimming when there is not enough oxygen. The mixture of LCLC with dormant bacteria behaves as a regular equilibrium LC, with the uniform director $\hat{\mathbf{n}}_0 = (1,0,0)$, specified by the surface treatment in planar state, being the ground state. The immobilized bacteria align with their elongated bodies parallel to $\hat{\mathbf{n}}_0$, Fig.6a. In the experiment performed by Zhou et al [66**], the dormant bacteria were transformed into active swimmers by an influx of oxygen supplied from the side of the cell. Once the bacteria start swimming, they cause bend instability of the director, with the periodic deviation of the actual director $\hat{\mathbf{n}}$ from $\hat{\mathbf{n}}_0$, Fig.6b,c. The period $\xi$ of stripes decreases with the increase of activity. As the activity increases further, the stripe pattern becomes unstable against nucleating pairs of of ±½ disclinations, Fig. 6d. The dynamic pattern of nucleating and annihilating disclinations, Fig.6e, makes the director globally isotropic in the plane of the film, with the aligning action of the substrates getting lost. The average distance $\xi_d$ between the disclinations is on the order of $\xi$.

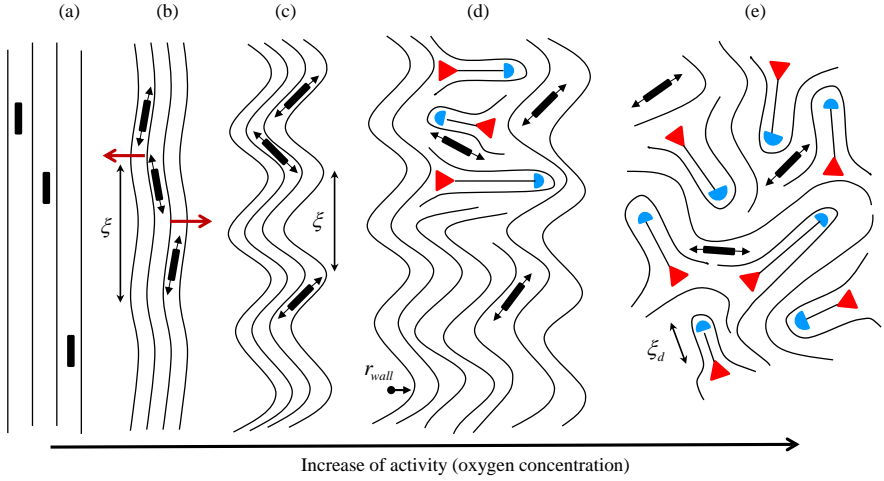

Increase of activity (oxygen concentration)

**Fig. 6.** *Development of nonequilibrium regimes in a living liquid crystal, following the experimental data in Ref. [66]. (a) Equilibrium state, dormant bacteria. (b,c) bend stripes instability, with the period $\xi$ decreasing as activity increases; (d) nucleation of $\pm 1/2$ disclination pairs; (e) topological turbulence. Swimming bacteria of pusher type are shown as rods with two arrows.*

The emergence of a stripe pattern is a general phenomenon expected for active matter in which the uniform orientational order is a subject of extension or compression and is thus unstable, see the review by Ramaswami [2]. In the case of living liquid crystals, the period $\xi$ can be derived from the balance of director-mediated elasticity and bacteria-generated flows. A swimming *B. subtilis* represents a pusher, a hydrodynamic force dipole of the strength (of a dimension of torque or energy) $U_0 \sim 1$ pN μm that produces two outward fluid streams coaxial with the bacterial body [90, 91]. A fluctuative realignment of the bacterial body away from $\hat{\mathbf{n}}_0$ would create a component of the fluid stream in the direction perpendicular to $\hat{\mathbf{n}}_0$ (shown by horizontal arrows in Fig.6b) and a reorienting hydrodynamic torque $\sim\alpha(h)cU_0\theta$, where $\alpha(h)$ is a dimensionless factor that describes the flow strength and depends on the cell thickness $h$; $c$ is the concentration of bacteria, and $\theta$ is the angular difference between local orientation of bacteria the director. The hydrodynamic torque is opposed by the restoring elastic torque $\sim K\dfrac{\partial^2 \theta}{\partial x^2}$; the balance defines a "bacterial coherence length" $\xi = \sqrt{K/\alpha cU_0}$; the model shows a good correspondence with the experimentally measured parameters [66**].

If the active component is of a "puller" type, one would expect an instability of the splay type [2]. Voituriez et al [92] considered an analog of the Frederics transition in a planar cell of an active polar film of a contractile (puller) type. If the activity is low, the film preserves its uniform static polar order; as the activity (or the cell thickness) increases above some threshold, the orientation in the center of the cell becomes progressively tilted, which implies splay, and the system flows. A homeotropic cell of an extensile (pusher) type should show a similar active Frederiks transition, of the bend type.

The increase of activity in the living liquid crystal thus shows a two-step scenario of instabilities: first, the uniform static director field transforms into a periodic bend stripes pattern, Fig.6b and second, the bend stripes, developing stronger director distortions of both bend and splay type, give rise to nucleating pairs of $\pm 1/2$ disclinations, Fig.6d. The latter help to randomize the system in the sense of director orientations: each disclination pair introduces a patch of the director that is aligned perpendicularly to the previously existing director, Fig.6d. The situation is similar to the Kosterlitz-Thouless model of phase transitions in equilibrium systems. Furthermore, the two types of deformations, stripes and disclinations, have a different dependency on the characteristic length scales. The elastic energy per unit area of the cell with well developed stripes (in which the tilt angle is no longer small, $\theta \approx 1$, Fig.6d) scales as $F_{wall} \approx K/\xi r_{wall}$, where $r_{wall}$ is the curvature radius at the wall separating "zigs" and "zags" of the periodically realigning director. In contrast, the energy per unit area of the cell filled with disclinations of strength $\pm 1/2$ scales as $F_{\pm 1/2} \approx K \ln(\xi_d / r_c)/\xi_d^2$, where $r_c$ is the core radius of the disclination; within the core, the scalar order parameter changes substantially (roughly speaking, the nematic "melts") [8]. When $r_{wall}$ approaches $r_c$ from above, the stripes start to resemble singular walls with the raising energy density. The elastic stress at the walls can be released by nucleating pairs of disclinations, since $F_{\pm 1/2}/F_{wall} = \xi r_{wall} \ln(\xi_d / r_c)/\xi_d^2$ become smaller than 1 when $r_{wall}$ decreases to about $\xi / \ln(\xi / r_c)$.

The turbulent patterns of disclinations at small Reynolds number is a general phenomenon in the active matter, observed previously in experiments with active gels of microtubules [93] and in numerical simulations of active nematics, see, for example, [94-97]. Since the living liquid crystals allow one to tune the activity in the experiments by controlling the flux of oxygen, further exploration of collective effects in them is clearly in order.

## 8. CONCLUSION

The last decades saw an important shift in the studies of LCs in hybrid systems, such as LC droplets in isotropic environment or colloidal inclusions in LCs [9, 10]. These studies explored mostly equilibrium states. Current research is increasingly focusing on dynamics of LCs and generally soft systems as evidenced by achievements in science of artificial and natural micro-swimmers and active matter. Active colloids in LCs represent a relatively new type of the active matter. However, it is already clear that the LC environment brings principally new mechanisms to the externally and internally driven dynamics.

In the case of electrically driven colloids, the LC, used as an anisotropic electrolyte or as an anisotropic dielectric, allows an effective pathway of the local energy input through the distortions of the orientation order. Since the source of activity is rooted in the properties of the medium, the principle lifts many limitations imposed by isotropic media on the properties of electrokinetically active interfaces and particles. For example, a glass sphere in an isotropic fluid cannot be driven by dielectric or electro-conductive mechanisms when acted upon by a uniform AC electric field. In a LC, such a motion is possible, enabled by anisotropy of either conductivity or permittivity. LC-enabled electrokinetics can transport particles of practically any type, solid or liquid, charged or not, dielectric or metallic, symmetric or asymmetric, polarizable or not. The LC director, serving as a guiding rail, can be predesigned to provide an additional degree of freedom in the dynamics of particles. Further exploration of LC-enabled electrokinetics will help to understand better the complex coupling of anisotropic surface interactions, bulk elasticity, viscous response, mechanisms of ionic transport in an anisotropic medium, which is a part of a more general area of charge transport in organic matter [98]. Many-particle systems, such as Quincke rotators moving through the LC host, can be further explored as an experimental realization of the Vicsek transition in dynamic systems, in which the interactions could be controlled by tuning the LC properties.

It is too early to speculate on concrete devices that might be built on the basis of electrically driven ACLCs. The director field of a LC, especially if pre-patterned in a desired configuration, can be used to guide the colloidal particles, either of a passive type in microfluidic settings, as recently reviewed by Sengupta, Herminghaus and Bahr [99*], or of the active type considered in this review. Conversion of the translational motion of Quincke rotators into a rotation of inclusions in the smectic host might be used in micromotors [60]. LC can certainly be of use as a medium for artificial or biological autonomous active colloids to impose the desired trajectory of motion. So far, the experiments are limited to bacteria, but there might be also a possibility to combine chemical-reaction powered particles with, for example, lyotropic water-based LCs. LCs as anisotropic electrolytes offer the following advantages over their isotropic counterparts. First, the LCs can be driven by an AC field that sustains steady flows. Second, the flows can be pre-designed through the director patterning. Third, if the LC is of a thermotropic hydrophobic type, it allows one to transport water droplets with dissolved chemicals. Fourth, the flows of LC and dispersed particles can be controlled through the temperature and frequency dependence of material parameters such as anisotropy of conductivity and dielectric permittivity. One can envision that these features might be in demand for future laboratory-on-a-chip devices, where steady flows of different geometries are required in precise microfluidic applications.

Exploration of internally driven ACLCs, such as bacteria, is also of a fundamental and applied interest. Biological systems are crowded and dynamic processes in them often occur at the background of some orientational order; thus the knowledge of how the out-of-equilibrium behavior is coupled to the medium anisotropy is of importance. Further studies might help to use LCs for fine control of bacterial activity and maybe even find ways to encourage the bacteria to perform useful mechanical work. The living liquid crystals with controlled activity represent a versatile model system to study the relationship between the macroscopic dynamic patterns and underlying microscopic interactions. One of the underexplored directions in the area of internally driven ACLCs is the study of autonomous artificial particles such as bimetallic nanorods, Janus spheres, powered by chemical reactions [98]; the LC director might provide a guiding directionality for these particles, as it does in the case of living bacteria.

The research presented in this review would be impossible without the contributions of I. Aranson, A. Jákli, Y.K. Kim, I. Lazo, Yu. Nastishin, V. Nazarenko, C. Peng, O. Pishnyak, S. Shiyanovskii, A. Sokolov, T. Turiv, D. Voloschenko, Q.H. Wei, J. Xiang, S. Zhou, to whom I am very thankful. The work is supported by NSF grants DMR-1507637, DMS-1434185, and DMR-1121288.

**REFERENCES AND RECOMMENDING READING**